\documentclass[11pt]{article}
\usepackage{geometry}               
\usepackage{graphicx}
\usepackage{amssymb}
\usepackage{epstopdf}
\title{
           Understanding Confinement From Deconfinement}
\author{M. Baker\\  {\it Dept of Physics, University of Washington} \\  {\it P. O. Box 351650, Seattle WA 98195, USA}}
\date{}                                     

\begin{document}
\maketitle
\begin{abstract}

We use effective magnetic $SU(N)$ pure gauge theory with cutoff $M$ 
and fixed gauge coupling $g_m$ to calculate non-perturbative 
magnetic properties of the deconfined phase 
of $SU(N)$ Yang-Mills theory.  We obtain the response to an external 
closed loop of electric current 
by reinterpreting and regulating the
calculation of the one loop effective potential in Yang-Mills theory.  
This effective potential gives rise to a color magnetic charge density,
the counterpart in the deconfined phase of color magnetic 
currents introduced in effective dual superconductor theories of the 
confined phase via magnetically charged Higgs fields.  The resulting 
spatial Wilson loop has area law behavior. Using values of $M $ 
and $g_m $ determined in the confined phase, we find
$SU(3)$ spatial string tensions compatible with lattice simulations in 
the temperature interval $1.5\,T_c\, < \,T \, < \,2.5 \,T_c$. Use of 
the effective theory to analyze experiments on heavy ion collisions 
will provide applications and further tests of these ideas.

\end{abstract}
\section{Introduction}

The confined phase of $SU(N)$ Yang-Mills theory can be described by an 
effective theory coupling
magnetic $SU(N)$ gauge potentials ${\bf C}_\mu$ to three 
adjoint representation Higgs fields  \cite{BBZ1991}.  
The coupling of the potentials ${\bf C}_\mu$ to the magnetically
charged Higgs fields generate color magnetic currents which,
via a dual Meissner effect,
confine $Z_N$ electric flux to narrow tubes connecting a
quark-antiquark pair  \cite {Mandelstam}.
The dual gluon (quanta of the magnetic 
gauge theory) aquires a mass $M_g$.
For $SU(3)$, 
 $M_g  \,\sim \, 1.95 \sqrt {\sigma} $ \cite{BBZ1997}.
The effective theory is applicable at distances greater 
than the flux tube radius  $R_{FT} \sim \frac{1}{M_g} \sim 0.3 fm $.  
Since $SU(3)$ lattice simulations \cite{wenger2002} 
yield a  deconfinement temperature 
$T_c \,\approx \, 0.65 \sqrt { \sigma}$, the scale
$M_g \,\sim \, 3T_c$.  
   There is then a range of temperatures  within the interval $T_c\, < \, T\,< 3\,T_c$ where the effective theory 
should also be applicable in the deconfined phase.
We will use the theory in this temperature range to calculate spatial 
Wilson loops, quantities that are outside the perturbative realm of 
finite temperature Yang-Mills theory.

In Section 2 we review the use of the effective theory in the confined 
phase. In  Section 3 we point out that in the deconfined phase the 
Higgs fields, in contrast to the gauge potentials, do not form 
part of the massless sector of the theory.
We neglect them at temperatures not too close to $T_c$, so that the effective theory reduces to magnetic $SU(N)$ 
Yang-Mills theory with a
cutoff $M_g$ and gauge coupling constant $g_m $ fixed 
by fits of heavy quark potentials in the confined phase \cite{BBZ1997}.

In section 4 we show that the spatial Wilson loop of 
Yang-Mills theory is determined by the 
effective potential $U(C_0)$ of the magnetic theory in the background 
of a static dual scalar potential $C_0$.  We evaluate the 
one loop contribution to $U(C_0)$, and use it to calculate 
the spatial string tensions 
$\sigma_k (T)$ that measure the magnetic flux with $Z_N$ quantum 
number $k$ passing through a large loop. We find that
these string tensions are proportional to $k(N-k)$ (Casimir scaling), 
and that the predicted $SU(3)$ string tension is compatible with the
results of lattice simulations \cite{boyd1996} in the temperature range $1.5T_c <T< 2.5T_c$.

In section 5 we compare  $SU(N)$ lattice simulations of 
string tensions with lattice simulations \cite{north} of dual string 
tensions  $\tilde{\sigma}_k (T) $ (measuring $Z_N$ electric flux) 
in the temperature range $T_c < T < 4.5T_c$. 
We find that the temperature $T \sim 1.5\,T_c$ marks a "transition"
from a high temperature perturbative regime having
$\tilde{\sigma}_k (T) > \sigma_k (T)$ to a low temperature domain
where $\sigma_k (T) > \tilde{\sigma}_k (T)$.

In section 6  we compare the spatial string tension, 
calculated in the effective magnetic gauge theory,
with that  calculated \cite{cft}  in the large $N$, large 
't Hooft coupling limit of  $SU(N)$ $ \mathcal {N} =4$ super Yang-Mills
theory.

In the final section we summarize the results, discuss the
significance of this work and suggest extensions and further tests.

\section{Effective Theory of the Confined Phase}
\label{sec:confined}
   The effective theory describing the low energy excitations of 
SU(N) Yang-Mills theory is a long distance dual $SU(N)$ Yang-Mills
theory coupling non-Abelian magnetic SU(N) 
gauge potentials ${\bf C}_\mu$ to 3 scalar fields $\phi_i$, 
each in the adjoint representation of the magnetic gauge group. The Lagrangian $L_{eff}$  has the form
\cite{BBZ1991}
\begin{equation}
L_{eff} = 2 tr \left[-\frac{1}{4} {\bf G^{\mu \nu} G_{\mu \nu}} + \frac{1}{2} (D_\mu \phi_i)^2 \right] - V(\phi_i) \, ,
\label{Leff}
\end{equation}

where

\begin{equation}
{\bf G_{\mu \nu}} = \partial_\mu {\bf C_{\mu \nu}} - \partial_\nu {\bf C_\mu} - i g_m [ {\bf C_\mu, C_\nu}] \, , 
\label{gmunu}
\end{equation}
and
\begin{equation}
D_\mu {\phi_i} = \partial_\mu \phi_i - i g_m [{\bf C}_\mu, \phi_i] \, .
\label{dmuphi}
\end{equation}
$V(\phi_i)$ is a Higgs potential which has a minimum at nonzero values of $\phi_i$.
It is chosen so that the Lagrangian (\ref{Leff})
describes a dual superconductor on the border between type I and 
type II.

In the confined phase the magnetic gauge symmetry is completely 
broken via a dual Higgs mechanism in which all particles become 
massive.   (At least $3$ adjoint scalars are necessary to 
completely break the symmetry.)
The value $\phi_0 $ of the magnetic  Higgs condensate 
is determined by the location of the minimum in the Higgs potential, 
and the dual (magnetic) gluon acquires a mass 

\begin{equation}
M_g \sim g_m \phi_0 \,, 
\label{dualgluon}
\end{equation}
via the dual Higgs mechanism.

The simplest possibility for the vacuum condensate $\langle \phi_i \rangle \equiv \phi_{i0}$ has 
the color structure \cite{BBZ1991}.
\begin{equation}
\phi_{10}=\frac{\phi_0}{\sqrt{2N}} J_x \,, \, \phi_{20}=\frac{\phi_0}{\sqrt{2N}} J_y \,, \, \phi_{30}=\frac{\phi_0}{\sqrt{2N}} J_z \,, \,
\label{vacuum}
\end{equation}
where $J_x$, $J_y$, and $J_z$ are the three generators
of the N dimensional irreducible representation of the
three dimensional rotation group corresponding to angular momentum 
$J=\frac{N-1}{2}$.  Since any matrix which commutes with all three generators must be a multiple of the unit matrix, there is no $SU(N)$ transformation which 
leaves all three $\phi_i$ invariant and the dual gauge symmetry
is completely broken.

The excitations above the classical vacuum  of the effective theory are
flux tubes connecting a quark-antiquark pair
in which $Z_N$ electric flux is confined to narrow tubes
of radius $ \sim \frac{1}{M_g}$, at whose center  
 the Higgs condensate vanishes. 
Explicit solutions have been obtained  for $SU(3)$. 
The scale of the energy distribution in these electric flux tubes is determined
by the dual gluon mass $M_g$.
  Since  the effective theory describes 
fluctuations only at energy scales less than $M_g$, 
there is no physical excitation with this mass.

The effective theory has two parameters; $g_m$ and $M_g$. Their 
values, $g_m \approx 3.91$ and $M_g \sim 800MeV$, were 
determined \cite {BBZ1997}
by comparing the predicted $SU(3)$ static heavy quark potential with 
lattice simulations.  For distances $R > 0.3fm$ the lattice potential 
is well represented by the sum of a term linear  in $R$ and a
$\frac{1}{R}$ term, $\frac{A_{lattice}}{R}$ \cite {kaczmarek+zantow}.
The value of  $g_m$ is obtained by writing the lattice $\frac{1}{R}$ 
potential  in  an effective Coulomb form:
\begin{equation}
\frac{A_{lattice}}{R}=-\frac{4}{3}\frac{\pi}{g_m^2 }\frac{1}{R } \, .
\label{Vc}
\end{equation}
 The RHS of (\ref{Vc}) is the potential obtained by coupling
magnetic gluons to a Dirac string connecting a quark-antiquark pair
with a strength $\frac{2\pi}{g_m}$, which is the perturbative result of the effective
theory. The coefficient of the linear potential is proportional to
$\frac{M_g^2}{{g_m}^2}$ and determines the value of $M_g$ in terms of $\sigma$ and $g_m$.

The spin dependent and
velocity dependent heavy quark potentials calculated
with the above values of $g_m$ and $M_g$ \cite{BBZ1997}
are compatible with results obtained
from $SU(3)$ lattice simulations \cite{bali}.
Furthermore, predicted energy distributions 
in electric flux tubes are compatible with lattice results
for these distributions for values of $R$ ranging 
from $1.0fm$ down to $0.25fm.$ \cite{green}.  

The long wavelength fluctuations of the axis of the electric
flux tubes give rise to an effective bosonic string 
theory governed by the Nambu-Goto action \cite{Baker+Steinke}. These fluctuations are the 
low energy excitations of the effective theory.  
The value of $g_m$ obtained from (\ref{Vc}) includes the 
energy,
$-\frac{\pi}{12R}$,
of the long wave length oscillations of the axis of the flux
tube \cite {Luscher1981}.
The value of $g_m \approx  3.91$ 
is close to $4$, so that the main contribution
to $g_m$ comes from renormalization due to string fluctuations.
Short distance fluctuations at energy scales greater than $M_g$ 
do not enter in the effective theory, and  $g_m$ is the coupling 
constant defined at the fixed scale $M_g$.

\section{The Effective Theory in the Deconfined Phase}
\label{sec:deconfined}

An approximate one loop calculation \cite{BBZ1988} of a finite temperature effective potential for the Higgs fields  yielded a potential whose minimum moved to $< \phi_i>\,=\,0$ at a temperature $T\,\sim\,\phi_0$. The deconfinement temperature $T_c$ is then on the order of  $\phi_0$.  
Above $T_c$ the Higgs condensate vanishes, so the magnetic gluon 
becomes massless.  However, since the deconfinement temperature for 
$SU(N)$ groups with $N\geq 3$ is first 
order \cite{wenger2002, wenger2005}, the Higgs particles remain 
massive in the deconfined phase. (This first order phase transition 
was not seen in the calculation \cite{BBZ1988} since it did not 
include the contribution of a cubic term in the Higgs potential.)

Since above $T_c$ the Higgs fields do not form part of the massless 
spectrum, we will neglect them in our treatment of the deconfined 
phase. Then in the leading long distance approximation the effective theory reduces to a pure SU(N) Yang-Mills 
theory of magnetic gauge 
potentials $ { \bf C}_\mu \equiv ({\bf C}_0, { \vec  {\bf C} })$. 
Although inclusion of the Higgs fields is essential in order to 
describe the transition to the confined phase, we will see a signal
for this transition in the behavior of the pure 
gauge effective theory as the temperature is lowered toward $T_c$.

This theory has the same form as the microscopic 
electric theory, but with a fixed gauge coupling constant $g_m$ 
and  fixed ultraviolet cutoff $M_g$. The values of these two
parameters are determined by the effective theory description
of the confined phase. 
The magnetic gluons, which at $T=0$ confine $Z_N$ 
electric flux,  become the physical degrees of freedom of the effective 
theory at $T\, >\, T_c$. These quanta are "strongly" interacting ($g_m \approx 3.91$), but 
their interaction is cut off at distances  less than $0.3fm$. 
Because of the duality between the microscopic electric 
$SU(N)$ Yang-Mills theory and the effective long distance magnetic 
$SU(N)$ gauge theory, perturbative calculations of electric 
quantities in the microscopic theory can be adapted to calculate magnetic quantities in the effective theory.

\section{The Spatial Wilson Loop Calculated in the Magnetic Theory.}
\label{sec:wilson}

To test the idea of using the effective theory to calculate
magnetic quantities in the deconfined phase
we calculate spatial Wilson loops measuring
magnetic flux with $Z_N$ quantum number $k$ passing through 
a loop $L$. (The spatial Wilson loop has area law behavior both above 
and below $T_c$.) 
The temporal Wilson loop of Yang-Mills theory determining the 
static heavy quark potential
is the partition function of the effective dual theory in the 
presence of a Dirac string connecting a static quark-antiquark 
pair \cite{BBZ1991}.
Similarly the spatial Wilson loop 
is the partition function of the effective dual theory in the 
presence of a current of $k$ quarks circulating
around the loop $L $ ($k$ closed Dirac strings).  This current is the source of a
color magnetic field $ { \vec {\bf B}_k} = {\bf G}_{0 k}$, the magnetic analogue of the color electric field 
$\vec {\bf E}$ generated in the confined phase by the Dirac string 
\cite{BBZ1991} :
\begin{equation}
{\vec {\bf B}} = - {\vec \nabla} {\bf C_0 } - i g_m  [{ \vec  {\bf C},  {\bf C_0}}] - \partial_t {\bf {\vec C}} \, .
\label{vecB}
\end{equation}

The  operator  creating the closed Dirac string
is a singular dual gauge transformation
which changes by a factor $e^{2\pi i \frac{k}{N}}$ when it encircles a 
curve linking the loop $L$. 
It is the dual  of the spatial 't Hooft loop operator that creates a 
closed line of magnetic flux along a loop $L$ in Yang-Mills 
theory \cite{thooft,philippe}. 
The effect of the dual 't Hooft operator is to add to ${\bf C}_0$ an external potential ${\bf C}_0^{Dirac}$ which is the magnetostatic scalar potential produced by $k$ circulating quark loops, 
each carrying a steady current $I = \frac{ 2 \pi /g _m}{1/T}$ and one 
unit of $Z_N$ charge.
 (The total color charge transported along the Dirac string is $2 \pi/ g_m$ and the total elapsed Euclidean time is $1/T$.) 

 The $j$th circulating quark gives a contribution to ${\bf C}_0^{Dirac}$  proportional to a diagonal matrix ${\bf Y}_1^j$ whose $j$th diagonal element  is equal to $- \frac{(N-1)}{N}$ and whose remaining $N-1$ elements are equal to $\frac{1}{N}$. The sum over the $k$ quarks gives \cite {korthalsaltes}
\begin{equation}
{\bf C}_0^{Dirac} ({\vec x})   = \frac {2 \pi T}{g_m}  \frac {\Omega_S ({\vec x})}{4 \pi } {\bf Y}_k \, ,
\label{c0dirac}
\end{equation}
where $\Omega_S ({\vec x})$ is the solid angle subtended 
at the point ${\vec x}$ by a surface $S$  bounded by the loop $L$,
and where ${\bf Y}_k \equiv \sum_{j = 1}^k {\bf Y}_1^j$ is a diagonal matrix having its first $k$ elements equal to $-\frac{(N-k)}{N}$ and its remaining $N-k$ elements equal to $\frac{k}{N}$. (The  $Z_N$ matrix $e^{2 \pi i {\bf Y}_k}  = e^{2 \pi i \frac{k}{N}}$ reflects the $Z_N$ charge $k$ carried by the $k$ circulating quark loops.)
The gradient of ${\bf C}_0^{Dirac} ({\vec x})$ contains a term
(a magnetic shell)
localized on the surface $S$ defining $\Omega_S$ which is
cancelled by a corresponding term in the action.
The regular part of $- \vec \nabla {\bf C}_0^{Dirac} ({\vec x})$ 
gives the Biot-Savart magnetic field 
${\vec {B}}_{BS}( \vec {x})$ of the current loop:
\begin{equation}
- \vec \nabla {\bf C}_0^{Dirac} ({\vec x}) =
\frac {2 \pi T}{g_m} \oint_{L} {\frac{d{\vec y}\times {(\vec {x}-\vec {y})}}{{|\vec {x} - \vec {y}|}^3}} {\bf Y}_k  \equiv
 \frac {2 \pi T}{g_m} {\vec {B}}_{BS}( \vec {x}) {\bf Y}_k \, ,
\label{biotsavart}
\end{equation}

The spatial Wilson loop calculated in the dual theory is the partition function $Z$ of the effective theory with ${\bf C}_0$ replaced by 
${\bf C}_0 ^{Dirac} + {\bf C}_0$, 
divided by the partition function with   
${\bf C}_0 ^{Dirac} =0$. 
Because $\vec \nabla {\bf C}_0 ^{Dirac}$ is singular on $L$, the functional integration 
defining $Z$ is restricted to those configurations ${\bf C}_0 ({\vec x})$ which vanish 
on the loop. (This boundary condition 
eliminates the singular cross term 
$(\vec \nabla {\bf C}_0 ^{Dirac} + \vec \nabla {\bf C}_0)^2$ 
in the classical contribution to the action.)

\subsection{The Effective Potential $U({\bf C}_0)$} 

To  evaluate the partition function of the  effective theory
 in the deconfined phase, 
where there is no classical potential, 
requires calculating the one loop effective potential
$U({\bf C}_0)$ 
in the background of a static magnetic scalar 
potential ${\bf C}_0$:

\begin{equation}
e^{-\int {d{ \vec x} \frac {U({\bf C}_0)}{T}}} \equiv 
e^{-S^{1-loop}({\bf C}_0)} 
=  Det \left (- D^2_{adj} ({\bf C}_0) \right) \, .
\label{Z1loop}
\end{equation}

We have calculated 
$U({\bf C}_0)$ integrating over the massless gauge modes of the 
magnetic theory and introducing a Pauli-Villars regulator mass $M$ to 
account for the short distance cutoff of the dual theory. 
This regulator mass $M$ should then be approximately equal to the 
dual gluon mass $M_g$ determining the maximum energy of the modes
included in the effective theory.
The calculation of $U({\bf C}_0)$, aside from the presence of the 
regulator, mimics the calculation of the one loop effective potential 
$U({\bf A}_0)$ in Yang-Mills theory \cite{gross, gocksch} used to 
evaluate the spatial 't Hooft loop \cite{gocksch, kovner} . We assume that the background potential 
${\bf C}_0$ has the same color structure 
as ${\bf C}_0^{Dirac}$, i.e.,
${\bf C}_0  =  \frac {2 \pi T}{g_m} C_0  (\vec x)  {\bf Y}_k$ . 
The corresponding effective potential $U(C_0)$ 
 is a periodic 
function of $C_0$ 
with period $1$ , having minima 
at the inequivalent $Z_N$ vacua of the magnetic theory.  

The result for the 1-loop effective action $S^{1-loop} (C_0)$ is

\begin{equation}
S^{1-loop}( C_0)  = \frac {k (N-k) (2 \pi T)^2 T^2}{3g_m^2} \int {d{\vec x} \frac {U( C_0)}{T}} \, ,
\label{S1loop}
\end{equation}

where

\begin{equation}
 U (C_0) =  \left [ {[C_0 ]}^2(1 - [C_0 ])^2  - \frac{3}{4 \pi^4} I ( C_0, \frac{T}{M}) \right]  \, ,
\label{U}
\end{equation}

and

\begin {equation}
I ( C_0, \frac{T}{M})  = \int_0^\infty { dy \,\,y^2 log \left (\frac{cosh \sqrt{y^2 + (\frac{M}{T})^2} - cos (2 \pi C_0)}{cosh \sqrt{y^2 + (\frac{M}{T})^2} -1 } \right)}  \, ,
\label{IC0}
\end{equation}
with $[C_0 ] \equiv {|C_0 |}_{mod1}$. The factor $2k(N-k)$ is the number of non zero eigenvalues of 
the matrix $\bf {Y}_k$ in the adjoint representation \cite {korthalsaltes,altes04}. 
The integral $I (C_0, \frac{T}{M})$ reflects the presence of the Pauli-Villars regulator in the functional determinant  (\ref{Z1loop}),  which
 suppresses the short distance contribution to
 $U(C_0)$. 
The one loop effective potential is ultraviolet finite, so that
in the absence of a regulator ($M \rightarrow \infty$),  
$I \rightarrow 0$. In this limit the one loop
 effective potential (\ref{U}),
 with $C_0$ replaced by $A_0$ and 
$g_m$ replaced by the running Yang-Mills coupling constant $g(T)$,
reduces to $U(A_0)$. 

The one loop  expression for $U(A_0)$ is applicable only at high 
temperatures where $g(T) \rightarrow 0 $ so that the contribution of higher order 
loops are small. The effective theory, by contrast, contains only modes
having energies less than the mass scale $M$. We will see that,
for temperatures greater than $T_c$, the one loop effective potential
$U(C_0)$ generates a classical solution $C_0 (\vec x)$ having a
mass scale $m_{mag} (T)$ greater than $M$. Consequently, in the 
long distance effective theory, higher order corrections in the loop
expansion about the classical solution $C_0 (\vec x)$ can be neglected.
We can then use the one loop effective potential $U(C_0)$ to calculate 
Wilson loops  via the effective theory (even though the coupling 
constant $g_m$ is not small).

Replacing $C_0$ by $C_0 + \frac {\Omega_S}{4 \pi}$ 
in $S^{1-loop}$  (\ref{S1loop})
to account for the coupling to the Dirac string 
and adding the classical action gives the effective action, $S_{eff} (C_0)$:

\begin{equation}
S_{eff}( C_0)  = \frac{ 4 \pi^2 T k (N-k)}{N g_m^2} \int {d {\vec x} \left [ (- \vec \nabla  C_0 + {\vec B}_{BS} ) ^2 + {U}( C_0 + \frac  {\Omega_S}{4 \pi} ) \frac{N g_m^2 T^2} {3} \right ]} \, .
\label{Seff}
\end{equation}
The background field is subject to the conditions 
$C_0 (\vec x) \rightarrow 0$ for $\vec x$ on $L$, and 
$C_0 (\vec{x}) \rightarrow - \frac {\Omega_S (\vec x)}{4 \pi } $ as $\vec x \rightarrow \infty$.
The latter condition means that that the total 
magnetic field $\vec B(\vec x) =\vec B_{BS} - \vec \nabla {C_0} $
is short range, decaying to its vacuum value at
large distances from the loop. 
$S_{eff} (L)$, the minimum value  of $S_{eff}(C_0)$,
determines the spatial Wilson loop, $e^{-S_{eff} (L)} $ , as calculated
in the effective theory. 

The term in (\ref{Seff}) linear in $\vec B_{BS}$ is a
surface term which gives no contribution to $S_{eff}$ because $C_0$
vanishes on $L$. The term quadratic in $\vec B_{BS}$
(the magnetic energy of a current loop of a thin wire) is proportional
to $L$ with a coefficient which diverges logarithmically as the 
thickness of the wire goes to zero. This ultraviolet divergence 
can be absorbed into a renormalization of the energy of the current 
source, after which the first term in (\ref{Seff}) becomes simply
$( \vec \nabla C_0)^2$ and only the second term in (\ref{Seff})
contains the external potential explicitly.

Because of the periodicity property of the effective potential,  $U( C_0) = U(C_0 + 1)$, 
the value of $U( C_0 + \frac { \Omega_S}{4 \pi})$ is independent of the choice of the surface $S$
defining the  solid angle $\Omega_S (\vec  x)$, and
we can choose $S$ to be the plane surface bounded the loop $L$.
For a square loop of side $L$ in the $xy$ plane centered
at the origin 
\begin{equation}
\Omega_S (x,y,z)  = - \int_{-\frac{L}{2}}^{\frac{L}{2}} dx^{\prime}
 \int_{-\frac{L}{2}}^{\frac{L}{2}} dy^{\prime}
\frac{z}{ [(x-x^{\prime})^2 + (y-y^{\prime})^2 +z^2]^{3/2}} \, .
\label {omega}
\end{equation}
Since $U(-C_0) = U(C_0)$ and $\Omega_S (x,y,-z) = - \Omega_S (x,y,z)$,
in minimizing (\ref{Seff})  we can consider configurations $C_0 (x,y,z)$
which are odd functions of $z$ 
so that  $C_0 = 0$ at $z=0$ for all $x$ and $y$ and the boundary 
condition on the loop is then automatically satisfied.

 \begin{figure}[ht] 
   \centering
    \includegraphics[width=5in]{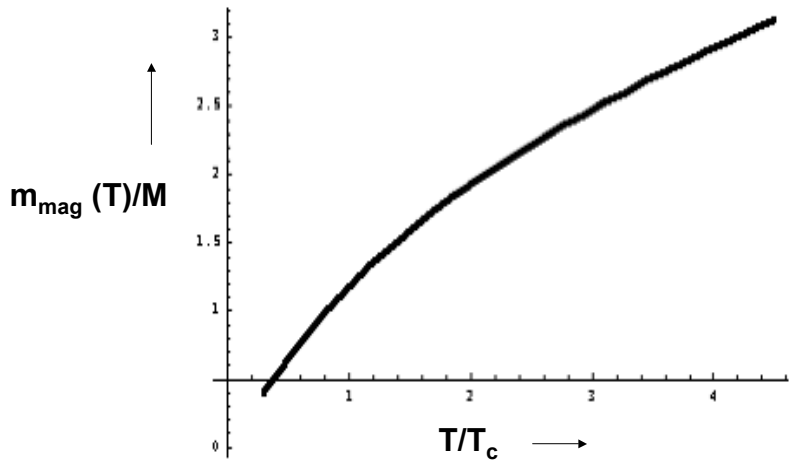} 
 \caption{ {\small Ratio of dual screening mass to the regulator
mass $M$ as a function of $\frac{T}{T_c}$ for $SU(3)$, with $T_c = \frac{M}{3}$ and $g_m\,=\,3.91$.  }}
   \label{fig:magneticmass}
\end{figure}

\subsection {The Magnetic Energy Density Profile $\vec{B}^2 (z)$}

The minimization of $S_{eff} (C_0)$  yields "Poisson's equation" for $C_0$:
\begin{equation}
- \nabla^2 C_0 (\vec x) = \rho_{mag} (\vec x) \, ,
\label{poisson}
\end{equation}
where
\begin{equation}
\rho_{mag} (\vec x) \equiv - \frac{1}{2} \frac{N {g_m}^2 T^2}{3}\frac{dU(C_0 + \frac{\Omega_S}{4 \pi} )}{d C_0}    
\label{magneticcharge}
\end{equation}
is the color magnetic charge density induced in the vacuum by the
current loop. This charge produces a field screening 
$\vec B_{BS}$, so that  the total field $\vec B (\vec x)$ 
has an exponential falloff determined by the dual screening mass 
$m_{mag} (T)$:
\begin{equation}
m_{mag} ^2(T) =  \frac{1}{2} \frac{d^2 U( C_0)}{d C_0^2} \bigg|_{C_0 = 0 }\left(\frac{N{g_m}^2 T^2}{3}\right ) \, .
\label{magneticmass}
\end{equation}

Using Eqs.(\ref{U}), (\ref{IC0}) and (\ref{magneticmass}),
in Fig. \ref{fig:magneticmass}  we plot
$\frac{m_{mag} (T)}{M}$ as a function of $\frac{T}{T_c}$ for $SU(3)$.
Note that for $T>T_c$, $m_{mag} (T) > M $, and that as the 
temperature is lowered toward $T_c$
the screening mass $m_{mag} (T)$, generated  from the fluctuations
of the massless quanta of the effective theory in the 
deconfined phase, decreases toward a value close to
$M$.

\begin{figure}[ht] 
   \centering
      \includegraphics[width=4in]{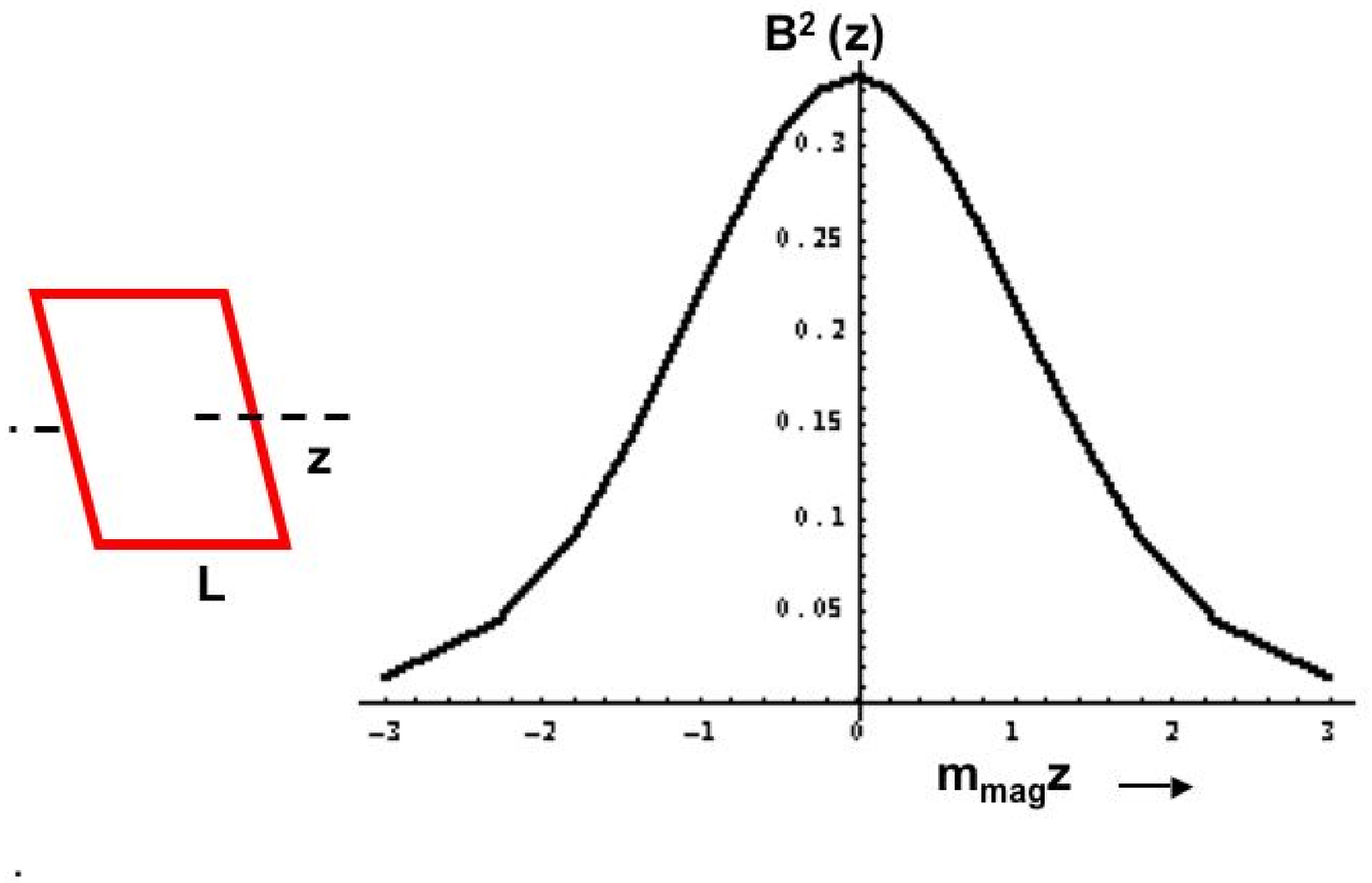}

   \caption{{\small Magnetic energy density profile $\vec{B}^2 (z)$  at $T = T_c$ as a function of  distance $z$ from $L$, calculated from  (\ref{poisson}).}}
   \label{fig:bsquared}
\end{figure}

The dual screening mass  $m_{mag} (T)$ determines the width of 
the magnetic energy profile surrounding a large spatial 
Wilson loop in the deconfined phase.  To find this profile we first 
take the limit $L \rightarrow \infty $ in (\ref{poisson}) and 
(\ref{magneticcharge}). In this limit $C_0$ and $\vec B$ are functions
only of the distance $z$ from the loop.
Furthermore the solid angle $\Omega_S = -2 \pi$  for $z>0$ 
and $ 2 \pi$ for $z<0$, so that the boundary condition at large distances 
becomes $C_0 (z) \rightarrow \pm \frac{1}{2}$ as $z \rightarrow \pm \infty$.

In Fig. \ref{fig:bsquared}  we plot $\vec{B}^2 (z)$  at $T = \frac{M}{3}$, obtained by solving (\ref{poisson}) with these boundary conditions.
Since for $T \,>\, T_c $ the mass $m_{mag} (T) $, which determines
the scale of the classical solution $\vec B (z)$,
is greater than the cutoff $M$ of the effective theory, 
then  (as pointed out in Sec. (4.1))  higher loop corrections  to the one loop effective potential
can be neglected.
In other words,  there are no small scale fluctuations
present in  the effective theory
to disturb the large scale structure of the classical solution,
and the one loop profile function $\vec B (z)$ is self-consistent.  

However, as $T$ is lowered to temperatures below $\frac{M}{3}$, where the 
width $\frac{1}{m_{mag} (T)}$ of the magnetic energy profile becomes 
larger 
than the minimum wavelength $\frac{1}{M}$ of the fluctuations included
in the effective theory,
the classical energy distribution is destabilized and the 
one loop approximation breaks down. 
The breakdown of the pure gauge effective theory at lower temperatures
is a signal for the transition to the confined phase 
for which the Higgs fields play an essential role.

\begin{figure}[ht]
   \centering
  \includegraphics[width=4in]{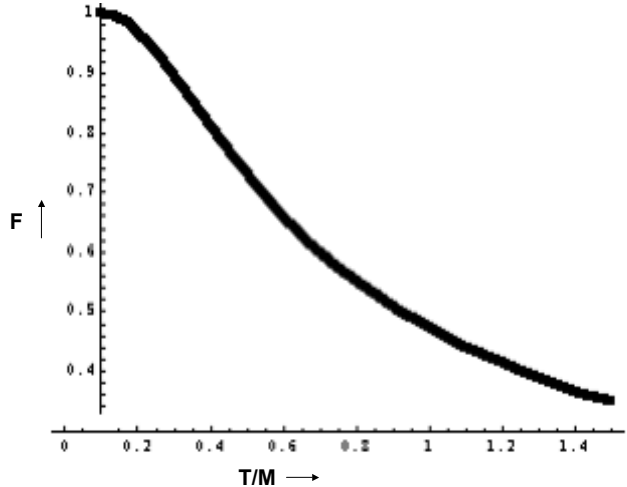}    
   \caption{{\small Function $F(\frac{T}{M})$,  defined in (\ref {functionF}), arising from a  Pauli-Villars regulator mass $M$, suppressing short distance contributions to the string tensions $\sigma_k (T) $.}}
   \label{fig:functionF}
\end{figure}

\subsection{Spatial String Tension:  Comparison with $SU(3)$ Lattice Simulations}

For large $L$ the effective action of the dual theory has area
law behavior determining the spatial string tension $\sigma_k (T)$:
\begin{equation}
S_{eff} (L) \rightarrow L^2 \sigma_k (T), \quad \mbox{as} \ L \rightarrow \infty \, .
\label{arealaw}
\end{equation}
Equivalently, the spatial string tension $\sigma_k (T)$ is the interface energy 
separating two vacua of magnetic $SU(N)$ gauge theory differing by
$k$ units of $Z_N$ charge \cite{north}.

The calculation of the spatial string tension follows closely the
corresponding calculation of the dual spatial string tension
$ \tilde{\sigma}_k (T)$, the interface energy 
in Yang-Mills theory \cite {gocksch, kovner}.
The one loop effective action evaluated at 
the "classical" solution $\vec B (z)$ 
yields: 
\begin{equation}
\frac{\sigma_k (T)}{T^2}=\frac{4\pi^2 k(N-k)F(\frac{T}{M})}{ 3g_m{\sqrt{3N}}} \, ,
\label{sigmakt}
\end{equation}
where
\begin{equation}
F(\frac{T}{M}) \equiv 6 \int_{- \frac{1}{2}}^{\frac{1}{2}} dC_0 
\sqrt {U(C_0 + \frac{1}{2})} \, .
\label{functionF}
\end{equation} 
Eq.(\ref{sigmakt}) is applicable for any $SU(N)$ group, but the values
of $g_m$ and $M$ have been determined only for $SU(3)$
where the effective theory has been applied in the confined
phase.
The function $F(\frac{T}{M})$, 
plotted in Fig. \ref{fig:functionF},
is the ratio of the action 
with regulator mass $M$ to the unregulated action. 
The temperature dependence of  the ratio $\frac{\sigma_k (T)}{T^2}$
comes from the Pauli-Villars cutoff, which suppresses the contributions 
of momenta greater than $M$ to $ \sigma_k (T) $. 
Since the Pauli-Villars regulator is rather "soft", 
allowing substantial contributions from momenta greater than $M$, 
we have also evaluated the string 
tension using values of $M$ smaller than $M_g \sim 800MeV$.

   \begin{figure}[ht] 
   \centering
     \includegraphics[width=5in]{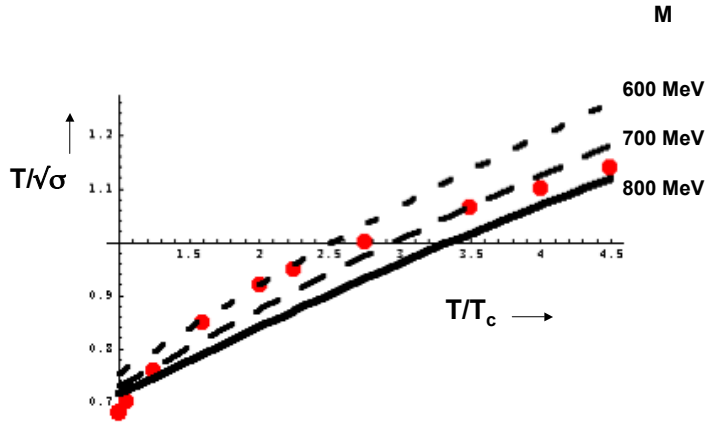} 
   \caption{{\small Comparison of $SU(3)$ 4d lattice data (dots) \cite{boyd1996,laine} for the spatial string tension $\sigma(T)$ with the prediction (\ref{sigmakt}) of the effective  magnetic Yang-Mills theory, for three values of the Pauli-Villars regulator mass $M$.}}   \label{fig:sigmakt}
\end{figure}
In Fig. \ref{fig:sigmakt} 
we plot $\frac{T}{\sqrt{\sigma (T)}}$ for $SU(3)$ ($k = 1, \sigma_k \equiv \sigma$) as a function of $\frac{T}{T_c}$  for Paul-Villars masses $M = 800 MeV$, $700 MeV$ and  $600 MeV$,
and compare with 
the results of 4d lattice simulations \cite{boyd1996,laine}.  
We note the following features of these curves:

\begin{itemize}

\item At $T \approx T_c$ 
the predicted values of 
 $\frac{T}{\sqrt{\sigma (T)}}$ 
lie close to the  lattice result, and they 
increase as the temperature increases, reflecting the
decrease with temperature of the function $F(\frac{T}{M})$ (\ref{functionF})
due to the Pauli-Villars regulator.

\item  $M = 600MeV$ gives the best fit 
to the $SU(3)$ lattice data in the temperature interval
$1.5T_c < T <  2.5T_c $ where the
effective theory should be applicable.

\item The value of the string tension does  not depend 
strongly on the Pauli-Villars mass.
(This reflects the ultra-violet 
finiteness of the one loop effective potential.) 

\item The lattice data in Fig. (\ref{fig:sigmakt}) are fit very well almost
down to $T_c$ by combining the non-perturbative value of the 
string tension of 3d $SU(3)$ Yang-Mills theory (determining the high 
temperature limit of the 4d string tension) with the 2-loop calulation 
of the running of the coupling constant $g_E (T)$ of 3d EQCD 
determining the change in the spatial string tension as the 
temperature is lowered) \cite{laine}.  
By contrast, the effective dual theory  determines the string 
tension in the deconfined phase only in a limited temperature range, 
but uses  parameters already determined in the confined  phase.  
The values of the intercepts of the curves in Fig. (\ref{fig:sigmakt}),
which are determined primarily by the value $g_m \approx 3.91$,
are predictions of the effective theory. For example, for 
$SU(8)$ and $k=1$, Eq. (\ref{sigmakt}) with $M \rightarrow  \infty$ 
gives $\frac{\sqrt{\sigma_1}}{T} \approx 1.72$,
while $SU(8)$ lattice simulations close to $T=T_c$ \cite{wenger2005}
give $\frac{\sqrt{\sigma_1}}{T} \approx 1.63$.

\end{itemize}

\subsection{Spatial String Tension $\sigma_k (T)$: Casimir Scaling}

We note from (\ref{sigmakt}) that $\sigma_k (T)$ is proportional 
to $k(N-k)$ (Casimir scaling).  This dependence on the quantum 
number $k$ of spatial string tensions in the deconfined phase is
consistent with the results of lattice simulations 
of $SU(4)$, $SU(6)$, and $SU(8)$ gauge 
theories \cite{wenger2005,korthalsaltes}.
Casimir scaling of the spatial string tension 
has also been obtained in a model of the deconfined phase
as a gas of non-abelian monopoles in the adjoint 
representation \cite{korthalsaltes,altes04}. 

On the other hand  such approximate Casimir scaling 
for the $T =0$ string tension would not be expected 
from the point of view of the effective theory because of
the presence of the Higgs condensate in the confined phase.

\begin{figure}[ht] 
\centering
\includegraphics[width=4in]{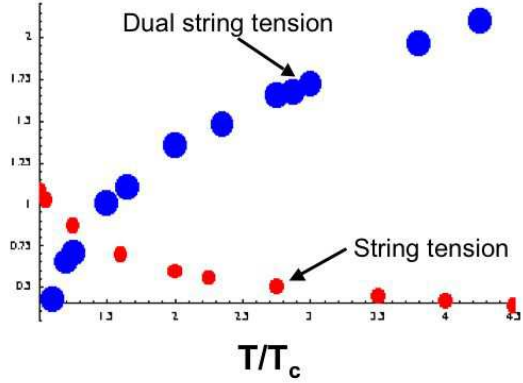}
\caption{{\small Comparison of dual string tension and string tension 
lattice data. Large dots: $SU(N)$ data ($N=3,4,6$ and $8$) for  
dual string tensions $\frac{ {\tilde{\sigma}}_k}{T^2}$,
divided by the Casimir factor $k(N-k)$, 
as a function of $\frac{T}{T_c}$  \cite{philippe}.
Small dots: Same plot of $SU(3)$ data for $\frac{\sigma}{2T^2}$  \cite{boyd1996}}.}
\label{fig:comparison}
\end{figure}

\subsection{Spatial String Tensions and Dual String Tensions Compared}

In Fig. \ref{fig:comparison} we compare the  $SU(3)$ lattice data  
for the string tension  with data for
dual string tensions $\tilde{\sigma}_k (T)$ 
measured in lattice simulations of $SU(3)$, 
$SU(4)$, $SU(6)$ and $SU(8)$ gauge theory in the temperature
range $T_c < T < 4.5 T_c $  \cite{north}. 
The lattice data for $\frac{\tilde{\sigma}_k (T)}{T^2}$ for all these 
$SU(N)$ groups and for all possible values of k, when scaled by 
the Casimir factor $k(N-k)$, all collapse on a single 
curve $\frac{\tilde{\sigma} (T)}{T^2}$  shown by the large dots 
in Fig. \ref{fig:comparison} .
This approximate Casimir scaling of dual string tensions 
agrees with the two loop perturbative prediction  
\cite{ korthalsaltes,gocksch}. 
At $T \approx 4.5 T_c$ the magnitude $\tilde{\sigma} (T)$ 
of the dual string tension agrees with the two loop 
perturbative prediction,  but at lower temperatures it
is suppressed \cite{north} relative to the perturbative prediction.  

This temperature range, where non-perturbative effects on
dual string tensions becomes significant, closely corresponds 
to the temperature range where the spatial string tension  
becomes comparable to the dual string tension. To show this, 
in Fig. \ref{fig:comparison} we also plot $\frac{\sigma (T)}{2T^2}$,
using the $SU(3)$ string tension lattice data in Fig. \ref{fig:sigmakt}.
We see that at $T\approx 4.5T_c$ the string 
tension $\sigma (T) \sim 0.2 \tilde{\sigma} (T) $ and
that, as the temperature decreases, $\frac{\sigma (T)}{T^2}$ 
increases, becoming greater than $\frac{\tilde{\sigma} (T)}{T^2}$ for 
temperatures less than $T \sim 1.25 T_c$.

We can then identify three temperature intervals in the deconfined
phase, each having distinctly different
electric and magnetic responses according to the value of the 
ratio $\gamma (T)$: 
\begin{equation}
\gamma (T) = \frac{\sigma_k (T)}{\tilde{\sigma}_k (T)} . 
\label{gamma}
\end{equation}

\begin{itemize}
\item  $ T > 4.5 T_c $,  $\gamma (T) < 1$: The dual string tension 
is perturbatively calculable, and the effective magnetic theory 
can not be used to calculate the string tension ($T\ge M$).  

\item $1.5T_c < T < 2.5 T_c $,  $\gamma (T) \sim 1$: 
The dual string tension is suppressed relative to its perturbative 
value, and the spatial string tension is calculable via
the effective magnetic theory.
($ T < M < m_{mag} (T) $). 

\item $T_c < T < 1.5 T_c $,  $\gamma (T) > 1 $: 
Neither perturbation theory nor
the effective magnetic theory are applicable.
In this temperature range $m_{mag} (T) \sim M $, which
is a signal for the transition to the confined phase.

\end{itemize}

\subsection{Spatial String Tension: Comparision with $\mathcal {N} = 4$ Super Yang-Mills Theory}

In this section we compare the  string tension predicted by
the effective magnetic theory with the expression $\sigma_{SYM}(T)$ for 
the spatial string tension  of $SU(N)$ 
$\mathcal {N} = 4$ super Yang-Mills theory, calculated in the large
$N$ limit and in the limit of large 't Hooft coupling 
$\lambda \equiv g_{SYM}^2 N $,
where the gravity-conformal field theory  correspondence is applicable \cite{cft}:
\begin {equation}
\sigma_{SYM}(T) = \frac{\pi}{2}\sqrt{\lambda}T^2  .
\label{SYM}
\end{equation}
There  is no scale in  $\mathcal {N} = 4$ SYM theory,  
$\lambda$ is a free parameter,
and the theory remains in the deconfined phase at all temperatures 
with $\frac{\sigma_{SYM}}{T^2} =  \frac{\pi}{2} \sqrt{\lambda}$.

In the scale free limit, $M\rightarrow \infty $,
the one loop result of the effective magnetic theory 
$\frac{\sigma_k (T)}{T^2}$ is also constant.
In this limit $F(\frac{T}{M})=1$ and (\ref{sigmakt}) becomes
\begin{equation}
\sigma_k (T)  =  \frac{4}{3\sqrt3}\frac{\pi}{2}\sqrt{\lambda_m} \frac{k(N-k)}{N} T^2 ,
\label{sigma11}
\end{equation}
where 
\begin{equation}
\lambda_m  \equiv (\frac{2 \pi}{g_m})^2 N 
\label{lambdam}
\end{equation}
is the 't Hooft coupling of the effective magnetic theory.
Eq. (\ref{sigma11}) is applicable
for any $SU(N)$,
but the value of $\lambda_m$ 
is known only for $SU(3)$ where $g_m = 3.91$  
gives $\sqrt{\lambda_m} =2.78$.

The factor $\sqrt{\lambda_m} $ in (\ref{sigma11}), 
determining $\sigma_k (T)$ in magnetic $SU(N)$ Yang-Mills theory,
is proportional to the width 
$\frac{1}{m_{mag} (T)}$ of the magnetic profile
multiplied by the number $N$
of unit $Z_N$ charges in the large $N$ limit.
The factor $\sqrt{\lambda}$ in (\ref{SYM}), 
determining $\sigma_{SYM}$,   
arises from the relation between the 't Hooft coupling and
the fundamental string scale via the
AdS/CFT correspondence.

The limit $N \rightarrow  \infty$ 
of (\ref{sigma11}) gives the factorized form:
\begin{equation}
\sigma_k (T) \rightarrow k\sigma_1 (T) =  k\frac{4}{3\sqrt3}\frac{\pi}{2}\sqrt{\lambda_m} T^2 . 
\label{sigma1}
\end{equation}
Since the string tensions $\sigma_{SYM} (T)$  and
$\sigma_1 (T)$ ((\ref {SYM}) and  (\ref{sigma1}))
have the same dependence on the 
't Hooft couplings of the two theories, 
the corresponding string tensions will be equal
if these two constants are related by a numerical factor
of order unity.
That is, imposing the relation
\begin{equation}
g_{SYM} =   \frac {4}{3 \sqrt{3}} \frac{2 \pi }{g_m} ,
\label{gcom}
\end{equation}
between the coupling constants $g_m$ and $g_{SYM}$
of the two theories, 
we obtain the equality of the two string tensions: 
\begin{equation}
\sigma_{SYM} (T) =  \sigma_1 (T) . 
\label{SYMcomparison}
\end{equation}
That is, with the correspondence (\ref{gcom}) the spatial string
tension $\sigma_{SYM} (T)$ is equal to the interface 
tension $\sigma_1 (T) $ of magnetic $SU(N)$ gauge theory 
calculated with the one loop effective potential. 
This correspondence provides a link between
effective magnetic Yang-Mills theory and $\mathcal {N} = 4 $
supersymmetric Yang-Mills theory.

\section {Summary}
\label{sec:summary}

We have used effective magnetic $SU(N)$ pure gauge theory
in the one loop approximation
to calculate spatial Wilson loops in the deconfined phase
in analogy to the use of the dual effective theory in the classical 
approximation to describe the confined phase.

Calculating the one loop effective potential for  {\bf$C_0$} with an ultraviolet cutoff  $M $, we find:

\begin{itemize}
\item At $T=\frac{M}{3} \sim T_c$  the width of the magnetic energy 
profile (Fig. \ref{fig:bsquared}) is approximately equal to the 
radius of the $T=0$ electric flux tube. 
\item In the temperature 
interval $1.5T_c<T<2.5T_c$ 
the predicted $SU(3)$ spatial string tension is compatible with 
lattice simulations (Fig. \ref{fig:sigmakt}). 
\item In this  temperature interval the values of the string tension 
and the dual string tension obtained from lattice simulations
(Fig. \ref{fig:comparison}) approach each other and become equal
as the temperature is lowered to about $1.25 \, T_c$. 
Roughly speaking, the temperature scale $M \sim 3T_c$ 
marks a "transition" in the behavior of the deconfined phase;  
the high temperature domain is described by perturbative  
Yang-Mills theory and the low temperature interval by the effective 
magnetic gauge theory.  
\item For SU(N) groups with $N \ge 3$ the string tensions 
$\sigma_k (T)$ satisfy Casimir scaling 
(while Casimir scaling is not expected in the confined phase).
\item With the duality correspondence (\ref{gcom}) the 
spatial string tension
$\sigma_{SYM} (T)$, calculated in $\mathcal {N} = 4$ SYM theory,
is equal to
string tension $\sigma_1 (T)$, calculated in the effective magnetic theory
in the scale free limit.
\end{itemize}

\section {Discussion}

The formation of the magnetic energy  profile
around a spatial Wilson loop in the deconfined phase parallels 
formation of an electric flux tube in the confined phase.

In the confined phase
an open Dirac string
connecting a quark-antiquark pair couples to the magnetic vector 
potential $\vec {\bf C}$ and induces a magnetic color
current density brought about by the 
interaction of the gauge potentials with the magnetically charged
Higgs fields.
Via the dual of Ampere's law this current density gives rise 
to an electric 
field $\vec {\bf E}=\vec \nabla \times \vec {\bf C}$ which 
screens the external Coulomb field generated by the open Dirac string,
so that the total color electric field decays exponentially
with the energy profile of an electric flux tube.

In the deconfined phase 
a closed Dirac string couples
to the magnetic scalar potential ${\bf C}_0$ and induces an effective
magnetic color charge density generated 
by the one loop effective $U({\bf C}_0)$.
Via the dual of Gauss's law this magnetic 
charge density gives rise to a magnetic  field
${\vec {\bf B}} = - {\vec \nabla} {\bf C_0 }$
which screens the external Biot-Savart magnetic field
 generated by the closed Dirac
string, so that the total magnetic field decays exponentially
at large distances and has the energy profile shown
in Fig. \ref{fig:bsquared}.
  
We thus gain an understanding of confinement by studying the
deconfined phase. 
The magnetic currents confining electric flux, introduced 
at the classical level via Higgs fields, are the counterparts in the 
confined phase of magnetic charges, generated in the deconfined phase
by integrating out the long distance quantum fluctuations of the
non-Abelian magnetic degrees of freedom.  
As the temperature is lowered
toward $T_c$ the confined magnetic energy profile resulting
from the one loop pure gauge effective action becomes unstable,
signaling the transition to the confined phase. Here
the inclusion in the effective action of the magnetically charged 
Higgs fields leads to topologically stable classical electric flux tube
solutions.

According to our picture, 
the effective theory describing the deconfined phase 
in a temperature range included in the interval $T_c\,<\,T\,<3T_c$ 
is $SU(N)$ Yang-Mills theory, just as in the microscopic theory. Only 
the physical interpretation of the potentials and the scale of the 
theory are altered. In the  temperature interval 
$1.5\,T_c\,<\,T\,<\,2.5\,T_c$ 
the deconfined phase 
consists of magnetic charges composed of massless magnetic gluons,
interacting "strongly" ($g_m \sim 3.91$) over distances
greater than $0.3fm$.  Since this temperature
interval is accessible in heavy ion collisions, 
calculations of non-equilibrium quantities in the effective theory
would make it possible to test the picture of 
the deconfined phase as a strongly interacting system
of magnetic gluons. Because of the presence of the long distance cutoff,
it should be possible to adapt perturbation
calculations, carried out in microscopic Yang-Mills theory
and applicable only at high temperatures, to the calculation of
corresponding properties in the effective magnetic Yang-Mills theory.

\section {Further Tests and Investigations}

\begin{itemize}

\item In the confined phase the long wavelength fluctuations of the 
axis of the flux tubes give rise to an effective bosonic string theory
and consequently to the $-\frac{\pi}{12R}$ L{\"u}scher correction
to the area law behavior of Wilson loops. In contrast, in the 
deconfined phase there is no Higgs condensate whose zeros locate 
the position of the string, and conseqently no effective string theory.
Instead, in order to calculate  the corrections to the area law behavior 
of spatial Wilson loops in the deconfined phase
in the temperature range where the effective theory is 
applicable we must solve  (\ref{poisson}) for 
finite values of $L$ and evaluate the corresponding effective 
action $S_{eff} (C_0)$ (\ref{Seff}). 
This calculation will be described in a separate paper.

\item Evidence for the magnetic quanta of the effective
theory should be sought in lattice simulations of Yang-Mills theory 
in the deconfined phase.

\item The effective magnetic Yang-Mills theory
should be used to analyze experiments on heavy ion collisions.
\end{itemize}

\section*{Acknowledgments}
I would like to thank  O. Aharony, B. Bringoltz, Ph. de Forcrand, M. Fromm, A. Karch,
 C. P. Korthals Altes, A. Vuorinen and L. Yaffe for their valuable help.

\end{document}